\documentclass[12pt]{iopart}
\usepackage{graphicx}
\usepackage[usenames]{color}
\begin{document}

\title[Customizing mesoscale self-assembly with 3D printing]{Customizing mesoscale self-assembly with 3D printing}

\author{M Poty$^{1}$, G Lumay$^{1}$ and N Vandewalle$^1$}

\address{$^1$ GRASP, Universit\'e de Li\`ege, B-4000 Li\`ege, Belgium.}

\begin{abstract}
Self-assembly due to capillary forces is a common method for generating 2D mesoscale structures from identical floating particles at the liquid-air interface. Designing building blocks to obtain a desired mesoscopic structure is a scientific challenge. We show herein that it is possible to shape the particles with a low cost 3D printer, for composing specific mesoscopic structures. Our method is based on the creation of capillary multipoles inducing either attractive or repulsive forces. Since capillary interactions can be downscaled, our method opens new ways to low cost microfabrication.
\end{abstract}

\maketitle

\section{Introduction}

\begin{figure}[h]
\begin{center}
\vskip 0.2 cm
\includegraphics[width=9.5cm]{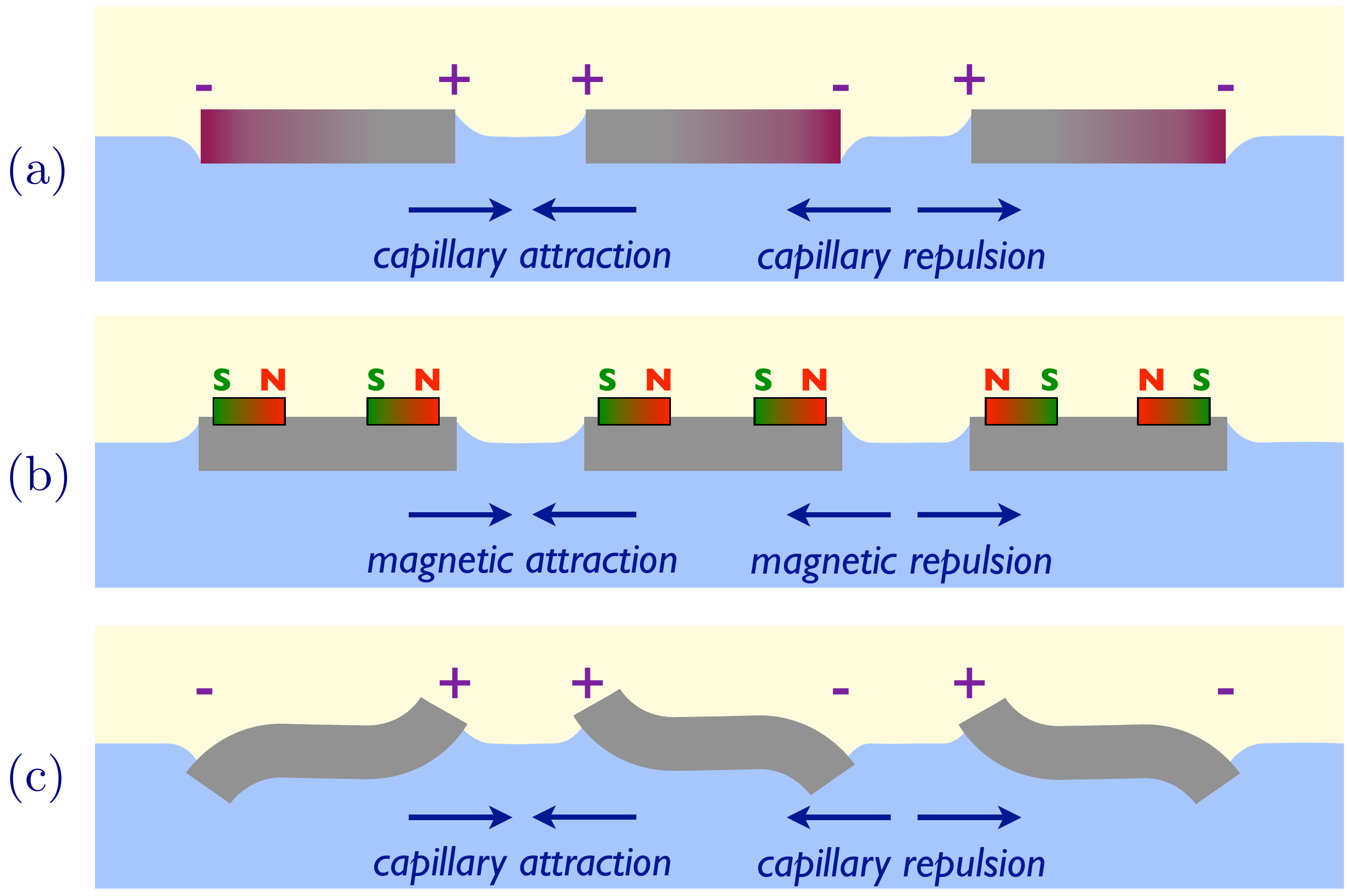}
\vskip -0.2 cm
\caption{Sketch of the interaction between neighboring particles. (a) The facets of particles can be treated differently in order to change the wetting properties : hydrophilic in grey, hydrophobic in red. This creates plus and minus capillary ``charges" inducing attraction and repulsion. (b) Incorporating magnetic dipoles into floating bodies, one obtains a magnetic force stronger than the capillary one. This interaction depends strongly on the dipole orientations. (c) The principle of our study is based on bent bodies, which induce local curvatures of the interface similar to (a). Since no material treatment is needed, those particles are easier to produce.}
\label{fig_interaction}
\end{center}
\end{figure}

Capillary driven self-assembly consists of suspending small objects at the water-air interface. Due to the balance between gravity and surface tension, the interface is slightly deformed, inducing a net force between the particles. Depending on the meniscus sign around each particle, both attractive or repulsive interactions can be obtained as illustrated in Figure \ref{fig_interaction}. Despite being a subject with tricky experiments, the fundamental and technological implications of the capillary effects are far from frivolous. Indeed, extensive researches demonstrated that the Ôself-assemblyÕ of small-scale structures can be achieved along liquid interfaces, opening ways to much simplified manufacture processes of micro-electromechanical systems \cite{selfassembly,pelesko}. 

A method \cite{bowden,bowden2} has been proposed to control the capillary interaction in order to obtain particular structures. By controlling the hydrophilic/hydrophobic character of the particle facets, one is able to obtain different capillary charges, as illustrated in Figure \ref{fig_interaction}(a). A modification of the density was also studied in order to enhance the interaction between particular objects \cite{bowden2}. Programming structures from the design of components has been envisaged \cite{pnas}. It has been also proposed \cite{golosovsky,pre} to incorporate magnetic dipoles into floating bodies (see Figure \ref{fig_interaction}(b)) for obtaining a magnetic force eventually stronger than the capillary one. It should be noted that this interaction depends strongly on the dipole orientations and it could be anisotropic along the liquid-air interface. Recently, dynamical features such as locomotion \cite{softmatter} and self-organized collective motions \cite{aranson}  have been obtained using oscillating external magnetic fields in such systems. 

In the present work, we propose an alternative method for controlling the liquid deformations around particles. The principle is illustrated in Figure \ref{fig_interaction}(c) : the particles are bent in such a way that their shape induces local curvatures of the liquid-air interface similar to Figure \ref{fig_interaction}(a), similarly to the behaviour of some water-walking insects \cite{insect}. Since no material treatment is needed, those particles are easy to produce. We will produce such floating objects using a low cost 3D printer in order to prove that our method can be implemented in every lab. 

\section{Experimental setup}

\begin{figure}[h]
\begin{center}
\vskip 0.2 cm
\includegraphics[width=15cm]{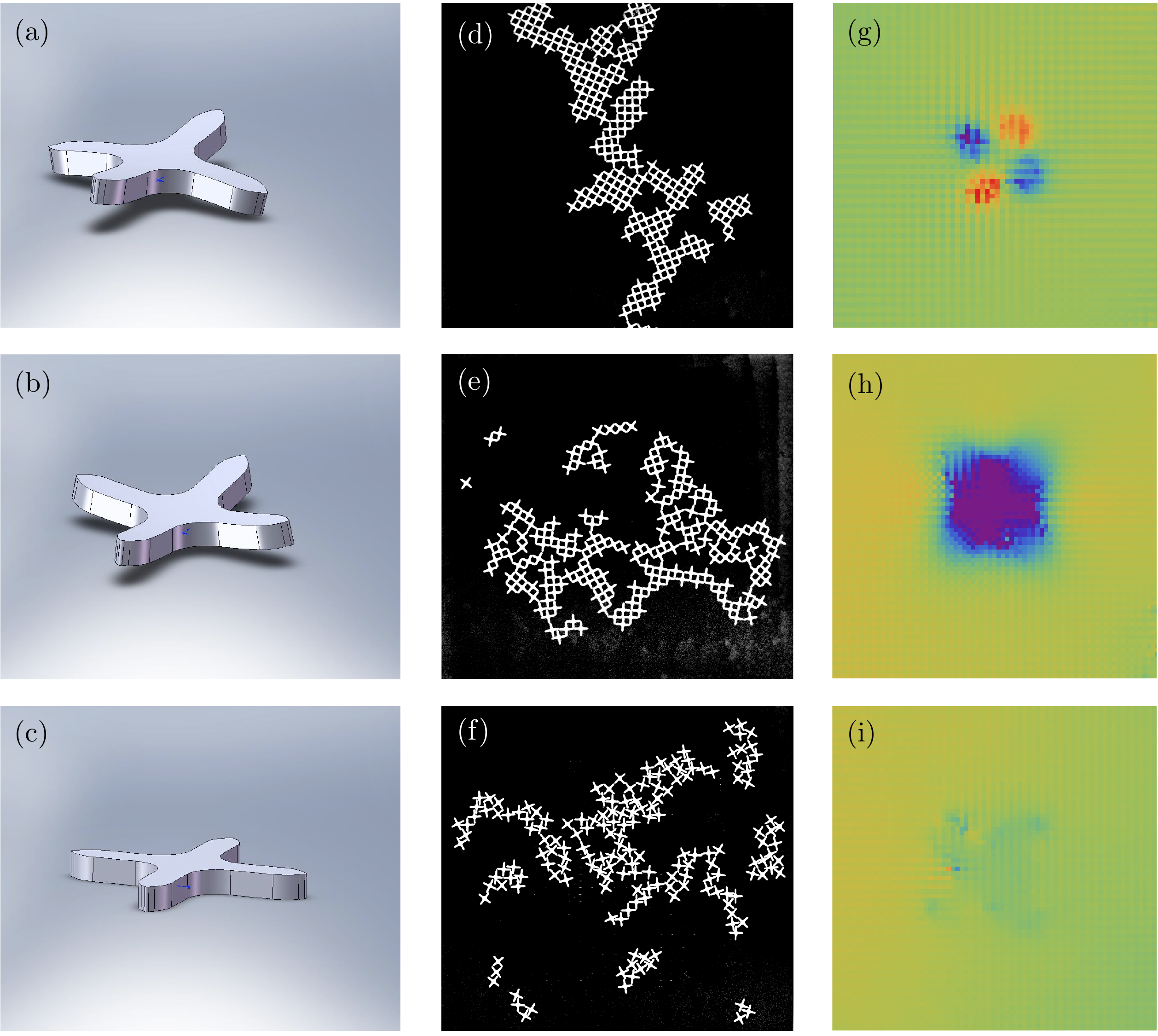}
\vskip -0.2 cm
\caption{Left column presents three shapes of floating bodies : (a) opposite curvature for successive branches and (b) branches with a similar positive curvature and (c) planar object as a reference. The central column shows top view of different assemblies (d,e,f) obtained with the shapes shown on the left. The scale is given by the size of the object $2b=15 \, {\rm mm}$. The right column shows three color scale images of the interface elevation around a single floating body. The different pictures correspond to the different printed objects of  : (g) opposite curvature for successive branches, (h) branches with similar curvatures, and (i) a planar object. } 
\label{fig_assembly}
\end{center}
\end{figure}

In addition to the above twist for controlling the sign of the capillary charge, the objects, that we will consider, are branched in order to create anisotropic interactions \cite{kralchevsky} that will be reflected in the subsequent symmetry of the self-assembled lattice. The branch number and lengths will control the target structure. We choose 4 branches of same length for creating square lattices. Other lattices or symmetries will be discussed at the end of this paper. Our capillary ``quadrupoles" are illustrated in Figure \ref{fig_assembly}. Three types of branches are considered in the present study : (a) curved objects with successive branches inducing positive and negative charges, (b) curved branches with four identical branches, and (c) non-curved objects. The branch length is fixed to $b=7.5 \, {\rm mm}$, such that the diameter is $2b$. The elements were produced using a  low cost 3D printer (Mojo from Stratasys). The apparent density of the Acrylonitrile Butadiene Styrene (ABS) material is 1.04. Partial wetting of the objects ensures their floatability at the interface.

Nearly two hundred elements are placed onto a metallic grid. The initial positions of the objects are assumed to be random and contacts are avoided. Then, the grid is slowly plunged into water : the grid sinks while the elements remain suspended at the air-liquid interface. Self-assembly takes place. After a few minutes, pictures are taken from above. This procedure is repeated many times for each system in order to obtain accurate statistics. The $40 \, {\rm cm} \times 40\, {\rm cm}$ container was specially built : a wedge has been created in order to pin the contact line with the container. By adjusting the liquid level with respect to the wedge vertical position, the boundaries can attract or repulse floating objects. In this work, we decide to select the liquid level which minimizes the boundary effects such that the forming self-assembly remains in the center of the container.  

\section{Results}

For planar objects, the liquid curvature near tips is similar to the curvature induced by the branches. Weak attractive interactions present in the system lead to either tip-tip contacts, tip-branch contacts or branch-branch contacts. The resulting pattern, seen in Figure \ref{fig_assembly}(f), is characterized by interlocked particles forming disordered structures. Since liquid deformations are very limited, as checked herebelow, weak capillary interactions are expected such that the structure is disconnected and seems composed of small clusters. 

When curved objects induce strong positive and negative capillary charges located near tips. The anisotropy of the interaction drives the self-assembly into regular square patterns, as shown in Figure \ref{fig_assembly}(d). Large crystal domains are seen. Similar structures are obtained with curved objects inducing only positive capillary charges, see Figure  \ref{fig_assembly}(e). However, the latter structures seem slightly less ordered than the previous ones. Indeed, crystal domains are smaller. The only difference between (d) and (e) cases is that repulsion exists in (d). 

\begin{figure}[h]
\begin{center}
\vskip 0.2 cm
\includegraphics[width=10cm]{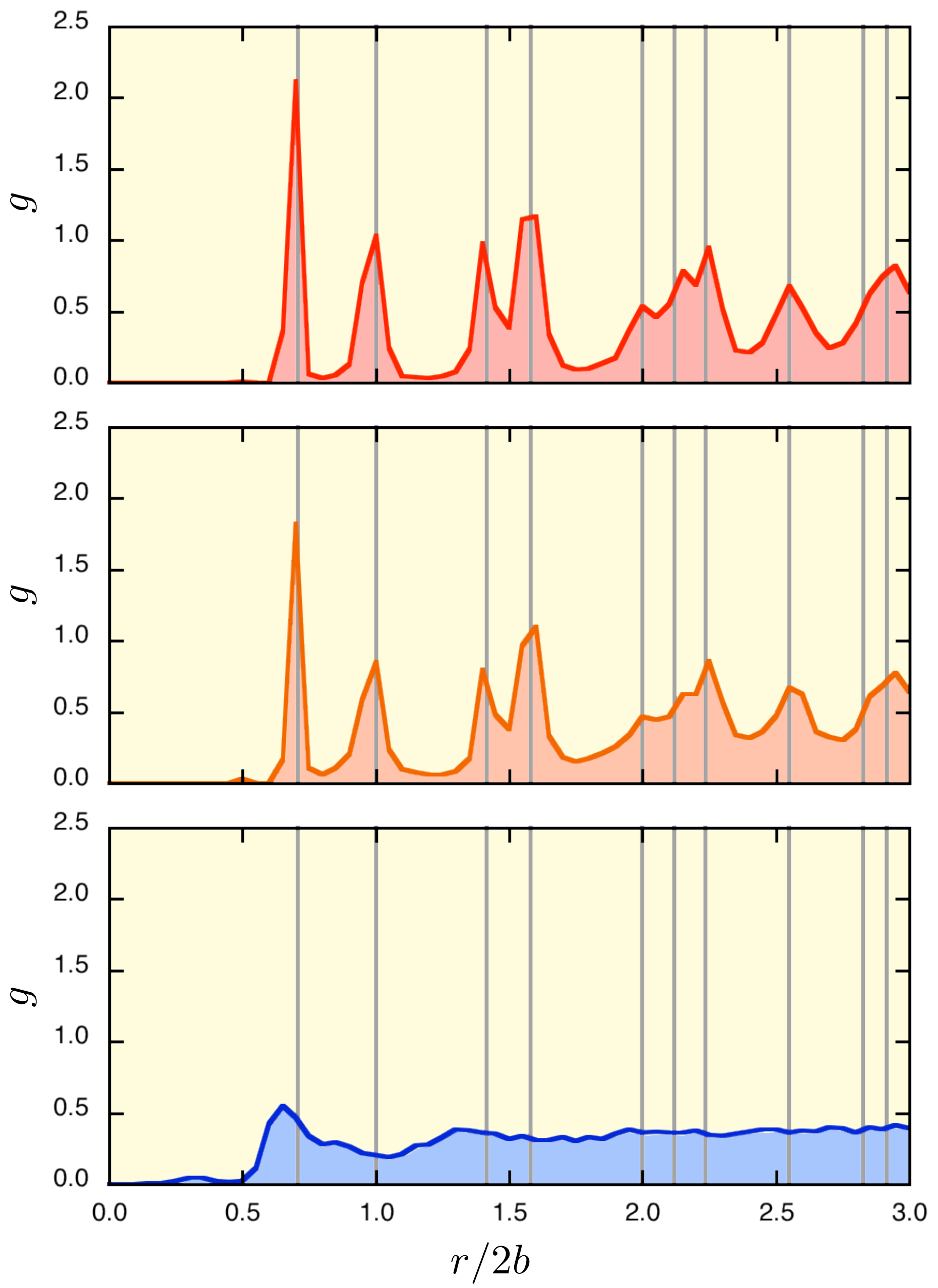}
\vskip -0.2 cm
\caption{Plot of the pair correlation function $g(r)$ for the three particles of Figure \ref{fig_assembly}. The horizontal axis is rescaled by the particle size $2b$ such that the nearest neighboring distance on a square lattice is $1/\sqrt{2}$. Vertical gray lines indicates the 10 first distances for a perfect square lattice. The curved objects show sharp peaks around expected distances. However, planar objects do not exhibit a specific order.  }
\label{fig_gr}
\end{center}
\end{figure}

By naked eye, one could experience some difficulties to distinguish between the patterns shown in Figure \ref{fig_assembly} (d) and (e). In order to measure the degree of ordering, we determine both position and orientation of each object on the liquid surface. We calculate the pair correlation function $g(r)$ for translational ordering. For each type of particle, the correlation function is shown in Figure \ref{fig_gr}. The distance $r$ is normalized by the object size $2b$ such that the neighbor distances are 
\begin{equation}
r/2b = 1/\sqrt{2}, 1, \sqrt{2}, \sqrt{5}/\sqrt{2}, ...
\end{equation}
for a perfect square lattice. It should be noted that the first distance is $1/\sqrt{2}$ because two nearest neighboring quadrupoles have ideally two tip/tip contacts. Vertical lines indicate the first 10 expected positions for the square lattice. The correlations function $g(r/2b)$ for curved objects show a series of well-defined peaks, indicating the formation of a square lattice. Nevertheless, the correlation function does not reach unity for large distances, meaning that the structure still contains voids at different scales \cite{arshad,dalbe}. It should be underlined that vibrating or rotating the interface can help the structure to reach denser structures \cite{rotating}. When positive and negative curvatures are considered (Figure \ref{fig_assembly}(a)), the first peak reaches about 2.1, being the mean number of nearest neighbors. In the case of identical curvatures (Figure \ref{fig_assembly}(b)), the first peak reaches a lower value around 1.8. For planar objects, a peak at $r/2b$ slightly below $1/\sqrt{2}$ is observed, indicating the occurence of tip-branch as well as branch-branch contacts in the structure. The peak expected for the second neighbors on a lattice is replaced by a valley and for larger distances, no particular structure is seen, meaning that the structure is highly disordered.

Ordered structures have been obtained for specific capillary quadrupoles. For characterizing our multipoles, we used an optical method proposed by Moisy et al.  \cite{moisy} for imaging the small deformation of a liquid-air interface around a single floatin body. The method is based on light refraction. A random pattern is placed at the bottom of the container, the camera records pictures of that pattern through a flat interface, i.e. without any object. After placing the object at the interface, new pictures of the pattern are taken. Series of pictures are compared and correlations are used to reconstruct the 3d shape of the interface. Figure \ref{fig_assembly} presents the resulting images for our three different floating bodies in the right column. The colors indicate the elevation of the interface around the object. Although the color scale is arbitrary, the same scale is used for the different patterns for comparison. Large deformations of the liquid interface are seen for curved objects (see Figure \ref{fig_assembly}(g) and (h)), while the planar body is seen to induce weak deformations (Figure \ref{fig_assembly}(i)). The curved objects induce quite different interface deformations : the object of Figure \ref{fig_assembly}(a) is seen to form positive (in red) and negative (in blue) lobes such that the attractive interactions are localized at the tips of the branches. The orientation of the object is therefore controlling the interaction. However, the printed object of Figure \ref{fig_assembly}(b) is forming a large depression around the object such that the object orientation is less relevant. One can understand that the alternation of negative and positive capillary charges around an object deeply affect the interactions between similar bodies, leading to highly elaborated lattices. 

\begin{figure}[h]
\begin{center}
\vskip 0.2 cm
\includegraphics[width=11cm]{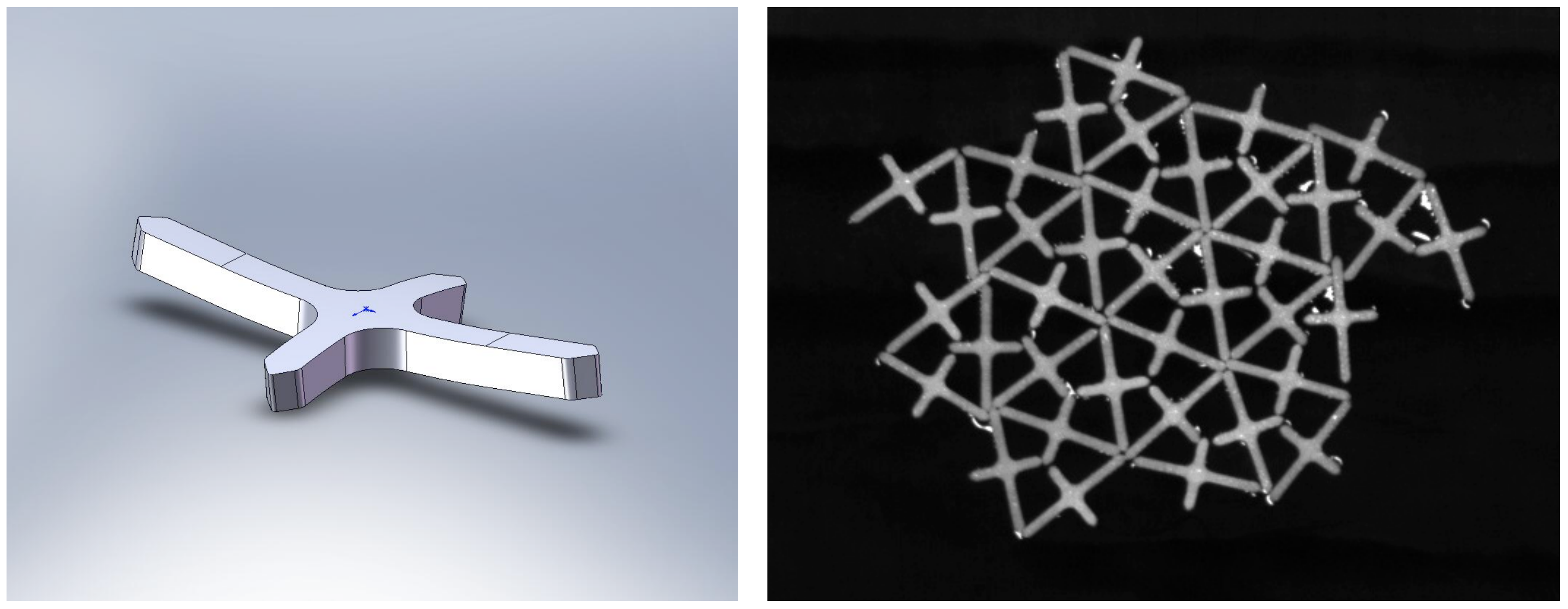}
\vskip -0.2 cm
\caption{(left) Image of the object with specific branch lenghts ($b_1/b_2= \sqrt{3}$) and opposite curvatures. (right) A triangular self-assembled structure is forming from such capillary quadrupoles. }
\label{fig_triangle}
\end{center}
\end{figure}

\begin{figure}[h]
\begin{center}
\vskip 0.2 cm
\includegraphics[height=4.7cm]{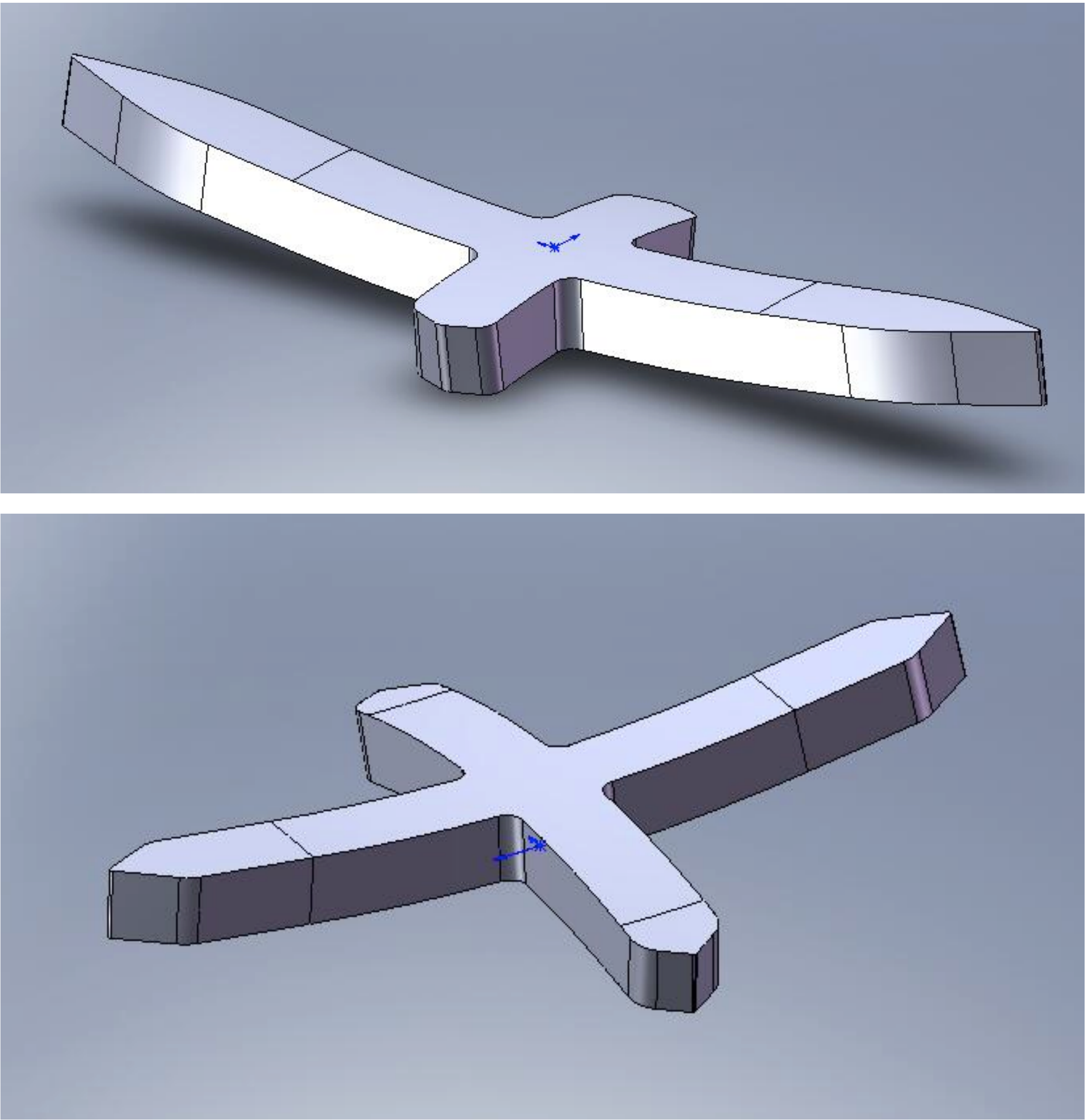}
\includegraphics[height=4.7cm]{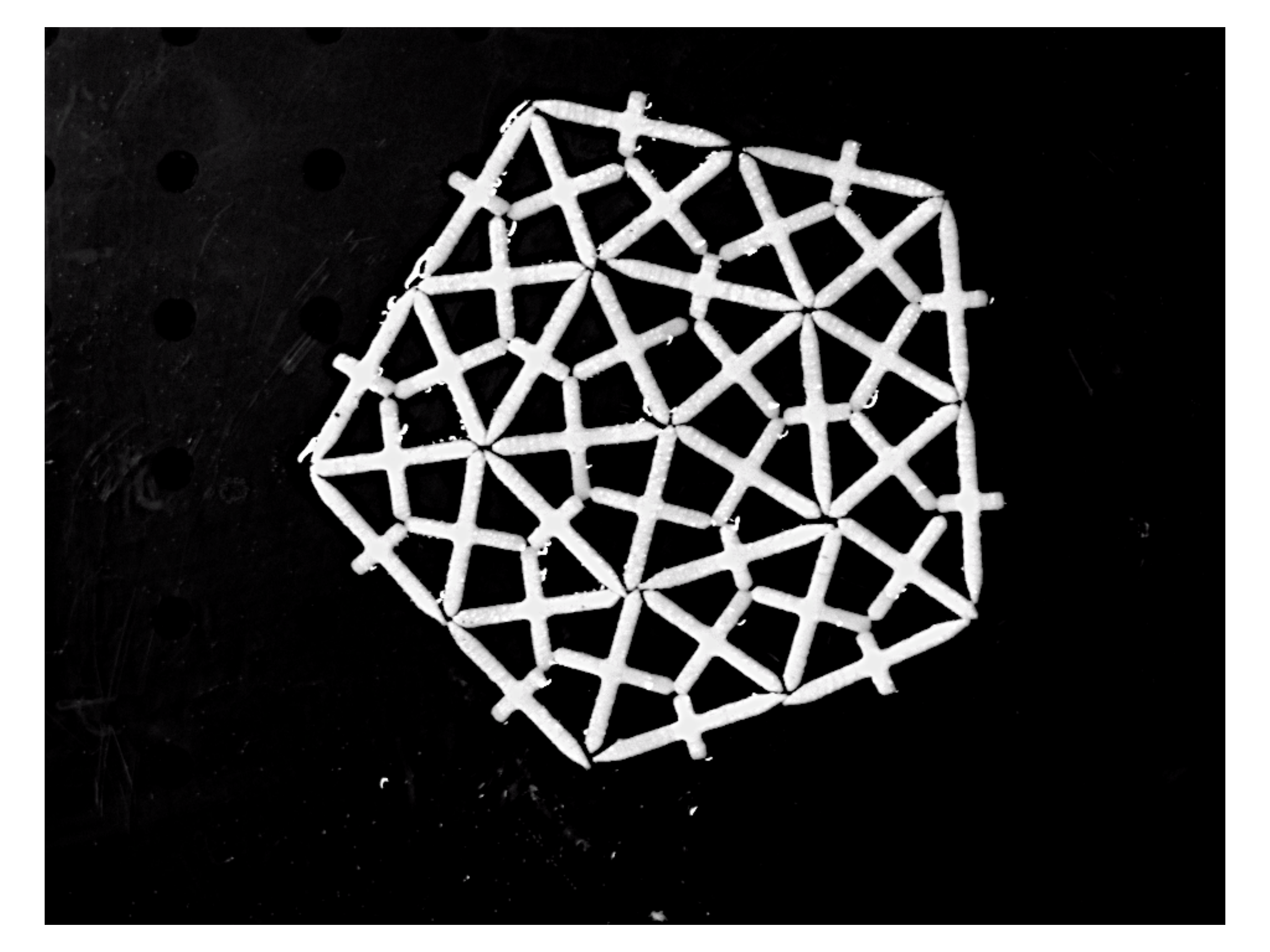}
\vskip -0.2 cm
\caption{(left) Images of two objects with specific branch lenghts ($b_1/b_2 = \tan(\pi/10)$ and $b_1/b_2 =\tan(\pi/5)$) and opposite curvature for enhanced capillary interactions. (right) A giant self-assembled pentagon made of 35 objects. In order to obtain this ideal pattern, the floating objects were launched successively along the interface. }
\label{fig_penta}
\end{center}
\end{figure}

Our study suggests that capillary multipoles are efficient building blocks for mesoscale directed self-assembly. In order to prove this concept, we created asymmetric crosses such that they combine into various structures. First, we created quadrupoles with non equal branch lengths :  branches are elongated in one direction. The branch length ratio is 
\begin{equation}
b_1/b_2= \sqrt{3}.
\end{equation} 
This object is shown in Figure \ref{fig_triangle}(left). Triangular lattices are formed from that elements as seen in the right picture of Figure \ref{fig_triangle}. It is possible to increase the complexity of the structure by combining two kinds of elongated objects. They are presented in the left column of Figure \ref{fig_penta}. They are characterized by different branch ratios : 
\begin{eqnarray}
b_1/b_2 = \tan(\pi/10), \\
b_1/b_2 =\tan(\pi/5).
\end{eqnarray}
The resulting pattern has a fivefold symmetry and is shown in Figure \ref{fig_penta}(right). This proves that complex structures can be achieved using compound systems. 

Since the capillary interactions can be downscaled to a few microns \cite{loudet,pre}, the actual developments of 3D printing towards small scales will allow scientists to imagine new ways to create mesoscopic self-assemblies without any specific surface treatment. Similarly to molecular recognition, capillary multipoles represent key ingredients for computing complex and highly elaborated mesostructures \cite{bonchevamrs}. 

\section{Summary}

In summary, we have proposed a convenient way to design particles for adjusting self-assembly  into desired mesoscopic structures. We investigate the large mesoscale structures containing up to hundred particles. We have shown that correlation functions are adapted for the characterization of self-assembled patterns. It should be underlined that even no mechanical agitation was used for improving the patterns, a remarkable degree of order has been reached. The study of interface vibrations will be studied in the future. 

This work opens new perspectives in self-assembly. Two subjects to be investigated are (i) active floating particles able to change their shape for changing/reseting the capillary interactions and (ii) the use of snowflake-like particles for designing a hierarchy of interactions allowing for complex tiling formation. In association with magnets \cite{bonchevapnas}, 3D self-assembly can also be envisaged.

\section*{Acknowledgements}
This work is financially supported by the University of Li\`ege (Grant FSRC-11/36).

\end{document}